\documentclass{iopart} 
\usepackage{amssymb,epsfig} 
\usepackage{psfrag} 
\usepackage{theorem}
\eqnobysec 
\def\th{\theta}
\def\ph{\varphi}
\def\la{\lambda}
\def\coupling{\sigma}
\def\kk{K_0}
\newcommand \del \partial 
\newcommand{\df}[2]{ \frac{\del {#1}}{\del {#2}} }
\newcommand{\ddf}[3]{ \frac{\del^2 {#1}}{\del {#2}\del{#3}} }
\newcommand \gt {\widetilde g}
\newcommand \Tt {\widetilde T}
\newcommand \Dt {\widetilde \nabla} 
\newcommand \Ocal {\mathcal O} 

\begin{document}

\title[Existence of naked singularities in Brans-Dicke theory]{Existence of
  naked singularities in Brans-Dicke theory of gravitation. An analytical and numerical study\footnote{Published in: {\tt Classical and Quantum Gravity (2010).}}} 
\author{ Nabil Bedjaoui$^1$,
  % Spelling:   LeFloch  or LeFLOCH
   Philippe G. LeFloch$^2$, \newline 
   Jos\'e M. Mart\'{\i}n-Garc\'{\i}a$^{3,4}$, 
   and J\'er\^ome Novak$^4$}
 \address{$^1$ LMFA \& INSSET, Universit\'e de Picardie Jules Verne, 48 rue
   Raspail, 02100 St. Quentin, France.}
 \address{$^2$ Laboratoire Jacques-Louis Lions \& Centre National de la
   Recherche Scientifique, Universit\'e Pierre et Marie Curie (Paris 6), 4 Place
   Jussieu, 75252 Paris, France. Blog: philippelefloch.wordpress.com.}
 \address{$^3$ Institut d'Astrophysique de Paris \& Centre National de la
   Recherche Scientifique, Universit\'e Pierre et
   Marie Curie, 98bis Boulevard Arago, 75014 Paris, France.}
 \address{$^4$ Laboratoire Univers et Th\'eories, Observatoire de Paris \& Centre National de la
   Recherche Scientifique,  Universit\'e Paris Diderot, 5 Place Jules Janssen, 92190 Meudon,
   France. 
   \newline\newline
   {\sl Email:} Bedjaoui@u-picardie.fr, pgLeFloch@gmail.com, 
   Garcia@iap.fr,  \newline and Jerome.Novak@obspm.fr.
}
\begin{abstract} Within the framework of the scalar-tensor models of gravitation 
and by relying on analytical and numerical techniques, we establish the existence of a 
class of spherically symmetric spacetimes containing a naked singularity. 
Our result relies on and extends a work by Christodoulou on the existence of naked singularities 
for the Einstein-scalar field equations. We establish that a key parameter in Christodoulou's
construction couples to the Brans-Dicke field and becomes a dynamical variable, which enlarges
and modifies the phase space of solutions. 
We recover analytically many properties first identified by Christodoulou, 
in particular the loss of regularity (especially at the center), and then 
investigate numerically the properties of these spacetimes. 
\end{abstract}

%===================================================================================================

\section{Introduction}
\label{section1}

The issue of the validity of the (weak version of the) cosmic censorship conjecture remains one of
the most important open problems in classical general relativity. Roughly
speaking, it says that physically admissible solutions to the Einstein equations should not contain
naked singularites, that is, all
singularities formed in physically reasonable scenarios of gravitational collapse should be 
surrounded by event horizons and, hence, cannot send signals to far observers at future null infinity. 
A precise formulation of the conjecture can be found in
\cite{Wald97, Christodoulou99b}, together with the properties required on a solution to qualify as a
physically ``reasonable'' process of singularity formation. These properties concern the smoothness
and genericity of the initial conditions, and demand that the matter model
undergoing collapse cannot form singularities of non-gravitational origin.

Even though the conjecture is still far from being proven in general, no definitive 
counter-example has been found so far, neither in numerical simulations nor
in analytical investigations. An important step forward
in this respect was the numerical analysis of the
threshold for black hole formation. After the pioneering work of Choptuik
\cite{Choptuik93} it has become clear that it is possible to form a naked
singularity by fine-tuning (smooth) initial conditions toward the vicinity of
the threshold of formation of arbitrarily small black holes. It turns out that the process is
dynamically controlled by an unstable exact solution --referred to as a
{\em critical solution}-- which, itself, does contain a naked singularity. The
fine-tuning is required to compensate for the instability of this solution and achieve a continuous
approach to that solution. In that set-up, there is no dynamical formation of naked
singularities and, therefore, this analysis does not provide a genuine counter-example to cosmic
censorship. (See \cite{LRR10} for a review.) 

In parallel to Choptuik's work on the critical collapse of a real massless
scalar field in spherical symmetry, Christodoulou \cite{Christodoulou94}
studied the Einstein-scalar field equations from a fully analytical point of view.
In a truly remarkable series of papers about the global dynamics of solutions to
this system, he constructed a family of exact solutions parametrized by some reals $k,a_1$ 
and showed that these spacetimes do contain a naked singularity in certain range of $a_1$, provided 
$0<k^2<1/3$. Later, in \cite{Christodoulou99}, he also established that these naked singularities are unstable
under small perturbation. Christodoulou's work provided the first complete mathematical proof of the formation of a naked singularity under gravitational collapse.

By construction, Christoudoulou's spacetimes are homothetic, that is, {\sl continuously self-similar} 
and, therefore, do not contain any privileged scale 
(such as a horizon of finite size), and so cannot contain a (finite) black hole.
Consequently, these spacetimes are a~priori good candidates to contain naked singularities 
with a central point of infinite curvature, denoted by $\Ocal$; see \cite{LakeZannias92}.
The critical solution found by Choptuik also possesses 
self-similarity, but of a different type, known as {\sl discrete self-similarity.} 
Since the symmetries are different, Christodoulou's solutions cannot ``relax''
to the critical solution, and actually the relation between the two solutions 
is unclear ---a problem that would deserve further study.

The parameter $k$, whose origin is in the massless scalar field (which only enters via its derivative 
in the Einstein equations), is essential in Christodoulou's construction, as well as the key 
restriction 
\begin{equation}
\label{123}
0<k^2<1/3.
\end{equation}
Depending on the second parameter $a_1$, an apparent horizon (rather than a naked singularity) 
is also possible in this range of $k$.
For 
\begin{equation}
\label{1234}
1/3 \le k^2 < 1
\end{equation}
 the future light-cone of $\Ocal$ collapses to a line, which provides an example of
a null singularity not preceded by an event horizon. For $k^2\ge 1$ the spacetimes are rather
pathological (see \cite{Brady95} and the Carter-Penrose diagram in Figure~4 of \cite{GM03}).

In all cases the parameter $k$ generates a mild loss of regularity at the center, which 
makes the curvature to be continuous but {\sl non-differentiable}
before the singularity at $\Ocal$ is formed. This implies that 
the past light cone of $\Ocal$ is non-regular, 
hence the initial conditions are not completely regular. 

Note that, on the contrary, Choptuik's critical solution is smooth everywhere except at the central
singularity at $\Ocal$ and, as a matter of fact, the sole requirements of regularity and discrete self-similarity
select a unique solution, at least locally in the phase space; see \cite{Gundlach95, MGGundlach03}.
If the massless scalar field is taken to be complex then it is possible to construct
a continuously self-similar solution which shares the regularity properties of the
Choptuik spacetime, and also contains a naked singularity, but is not critical
\cite{Hirschmann95}.

Our main objective in the present work is to investigate whether the relevance of $k$ 
and the role played by the limiting conditions on $k$ for the formation of naked singularities 
are ``structurally stable'', that is, whether spacetimes with the same features can be constructed
with extended models in which families of solutions with an
equivalent parameter are present. Indeed, the model we consider contains Christodoulou's
model as a special case. Specifically, we work here within the scalar-tensor theory of gravitation and, 
especially, within the so-called Brans-Dicke theory. 

The model under consideration here effectively adds an additional scalar field to
Christodoulou's system of equations, and makes $k$ a {\sl dynamical variable}, denoted by $K$.
Interestingly enough, our analysis leads to the same range in order to avoid the pathological behavior
referred to above, namely
\begin{equation}
\label{124}
0 < K_*^2 <1,
\end{equation}
where now $K_*$ is the value of the field $K$ at the past light cone of the singularity.
We also show that, starting from an arbitrary initial value for $K$, 
the system under consideration dynamically evolves toward values $K$ {\sl below} the threshold $1$.
Therefore, the introduction of extra degrees of freedom allows to avoid 
the pathological spacetimes arising with the Einstein-scalar system.

Note that Liebling and Choptuik \cite{Liebling96} have numerically shown the presence of critical
phenomena in the Brans-Dicke system, the critical solution being discretely or continuously
self-similar (depending a coupling constant). Again, being completely smooth, such a critical solution
is not related to the solutions we construct in this article. For further results, see [5,18,20,26]. 

The system under study is significantly more involved than the one studied analytically by 
Christodoulou \cite{Christodoulou94} and, although we do follow and generalize several important
steps in the construction therein, we eventually must resort to numerical investigations to reach
our final conclusions. In fact, by relying on numerics, 
we 	arrive at a better understanding of the class of solutions and are able to construct
explicit examples.  

An outline of this paper is as follows. 
In Section~\ref{section2}, we introduce the model of self-gravitating matter of interest,
and we determine the general evolution equations under the assumption of 
radial symmetry. In Section~\ref{section3}, we impose the self-similar assumption and 
show that general solutions are parameterized by four functions of a single variable, denoted by $x$, 
which obey a system of ordinary differential equations (ODE). 
We construct solutions that are piecewise regular, with each piece separated by
singular points across which careful matching is required.
Specifically, in Sections~\ref{section4} and~\ref{section5}, we successively construct the
interior and exterior part of the past light-cone of the singularity. 
Finally, in Sections~\ref{section6} and \ref{section7}
we describe our numerical strategy and present various results and conclusions.

%===================================================================================================

\section{Scalar-tensor theories}
\label{section2}

\subsection{Scalar-tensor gravity with scalar field}

Scalar-tensor theories of gravity are alternative theories of gravity
which are physically strongly motivated 
and have a long history in the literature. The fundamental assumption of these theories 
is that the gravitational field is mediated by one (or more) scalar field(s)
 in addition to the standard tensor field ($g_{\mu\nu}$) of
Einstein's general relativity. These theories satisfy the equivalence principle (since they are
metric-based theories), but do not satisfy the strong version of the equivalence principle. 
The first theory of this kind 
was developed by Jordan~\cite{Jordan59}, Fierz~\cite{Fierz56}, and Brans and 
Dicke~\cite{Brans61}, and contains an additional parameter defining
the coupling of the scalar field to the matter model. 
Later on,
Bergmann~\cite{Bergmann68}, Nordtvedt~\cite{Nordtvedt70}, and
Wagoner~\cite{Wagoner70} extended this approach to a coupling via
a function of the scalar field. Next, Damour and
Esposito-Far\`ese~\cite{Damour92} introduced a generalization based on an arbitrary number of scalar fields. 
More recently, cosmological models based on the
so-called $f(R)$ gravity theories have attracted a lot of attention, 
which found applications in the study of relativistic stars~\cite{Babichev09}. 
These theories form a subclass of the scalar-tensor theories, and it 
is interesting to look for a better understanding of the corresponding 
spacetimes and, in particular as we do in the present work,
to study the possible existence of spacetimes containing naked singularities. 

Specifically, we are going to investigate a generalization of Christodoulou's
model when a a scalar field $\phi$ in coupled to a scalar-tensor theory of
gravity for which the action reads (see \cite{Damour92,Santiago97} for
details):
\begin{eqnarray}
\nonumber 
  S &=& S_G + S_m
  \\ 
 &=&   \frac{1}{4} \int_M  \big( R -
    2g^{\mu\nu} \del_\mu \psi \del_\nu \psi \big) \, \sqrt{-g} \, \textrm{d}^4x 
    - \frac{1}{2}
  \int_M  \gt^{\mu\nu} \del_\mu
  \phi \del_\nu \phi \, \sqrt{-\gt} \, \textrm{d}^4 x. 
  \nonumber 
\end{eqnarray}
We use a system of units for the gravitational constant $G$ and the
light speed $c$ such that $\frac{4\pi G}{c^4}=1$.  The spacetime $M$ is
four-dimensional and is endowed with two conformally-related metrics: the
\textsl{Einstein metric\/} denoted by $g_{\mu\nu}$, and the
\textsl{Brans-Dicke (or physical) metric} denoted by 
$$ 
  \gt_{\mu\nu} = a^2(\psi) g_{\mu\nu}.  
$$
In the latter, $a(\psi)>0$ is a coupling function entering the theory, and one
recovers classical general relativity by choosing $a(\psi)$ to be constant.

This theory admits two scalar fields: one of them, $\phi$, 
represents the matter content of the spacetime and the other,
$\psi$, generates the gravitational field. 
When $a$ is not just a constant, the matter field $\phi$ does interact with the physical
metric $\gt_{\mu\nu}$, whereas the gravitational field equations for 
$g_{\mu\nu}, \psi$ are formulated in terms of $g_{\mu\nu}$, only.

%---------------------------------------------------------------------------------------------------

\subsection{Choice of coordinates}

Throughout this paper we use a notation consistent with the one in
Christodoulou~\cite{Christodoulou94}, in order to make easier the comparison between the two models.
We consider a general spherically symmetric spacetime whose metric is 
expressed in Bondi coordinates~\cite{Bondi62} as  
$$ 
  g= -e^{2\nu} du^2 - 2e^{\nu+\lambda} du dr + r^2 d\Omega^2, 
$$
in which the metric coefficients $\la, \nu$ depend on the coordinates $u,r$,
only, and $d\Omega^2$ represents the unit round metric on the 2-sphere.
The relevant components of the Ricci tensor $R_{rr}$
and $R_{\th\th}$  are found to be (see for instance~\cite{Bondi62,Puerrer07}): 
\begin{eqnarray}
\nonumber 
  & R_{rr} = \frac{2}{r}\left( \df{\la}{r} + \df{\nu}{r} \right),
  \\
  &  R_{\th\th} = r \left( \df{\la}{r} - \df{\nu}{r} \right) e^{-2\la} + 1 -
  e^{-2\la}, \nonumber
\end{eqnarray}
with $R_{\ph\ph} = (\sin\th)^2 \, R_{\th\th}$.
Observe that these formulas involve first-order derivatives
of the metric coefficients, only.
For completeness, we also determine the other two non-vanishing components of
the Ricci tensor, which now involve second-order derivatives of the metric, i.e. 
\begin{eqnarray}
\nonumber
  & R_{uu} = e^{2(\nu-\la)}\left( \frac{2}{r}\df{\nu}{r}
    -\df{\la}{r}\df{\nu}{r} +\left(\df{\nu}{r}\right)^2
    +\df{^2\nu}{r^2} \right)\nonumber
  \\ \nonumber
  & \qquad \quad
  - e^{\nu-\la} \left( \frac{2}{r}\df{\la}{u}
    +\ddf{\nu}{u}{r} +\ddf{\la}{u}{r}\right), \nonumber
  \\ \nonumber
  & R_{ur} = e^{\nu-\la}\left( \frac{2}{r}\df{\nu}{r} -\df{\la}{r}\df{\nu}{r}
    +\left(\df{\nu}{r}\right)^2
    +\df{^2\nu}{r^2} \right)
  -\ddf{\nu}{u}{r} -\ddf{\la}{u}{r}. \nonumber
\end{eqnarray}
Note in passing the following simple relation:
$$
  e^{\la-\nu} R_{uu} - R_{ur} = -\frac{2}{r}\df{\la}{u}.
$$

Alternatively, one may consider the Einstein tensor $ G_{\alpha\beta} =
R_{\alpha\beta} - \frac{1}{2} R\, g_{\alpha\beta}$ for the metric
$g_{\alpha\beta}$, and compute its essential components
\begin{eqnarray} \nonumber
&  G_{uu} = \frac{e^{2(\nu - \la)}}{r} \left( 2 \df{\la}{r} +
  \frac{e^{2\la}-1}{r} - 2e^{\la-\nu} \df{\la}{u} \right), \nonumber
\\
&  G_{ur} = \frac{e^{(\nu - \la)}}{r} \left( \frac{e^{2\la} - 1}{r} + 2
  \df{\la}{r} \right), \nonumber
\\
&  G_{rr} = \frac{2}{r} \left( \df{\nu}{r} + \df{\la}{r} \right).
 \nonumber
\end{eqnarray}
%  

%---------------------------------------------------------------------------------------------------

\subsection{Evolution equations}

By varying the action of the theory with respect to both
metrics, one gets two conformally-related stress-energy tensors
$T^{\alpha\beta}$ and $\Tt^{\alpha\beta} = a^{-6}(\psi) T^{\alpha\beta}$.
Denoting by $T=g_{\alpha\beta}T^{\alpha\beta}$ and $\Tt =
\gt_{\alpha\beta}\Tt^{\alpha\beta}$ their traces,  
one can check that (cf.~\cite{Damour92} for details): 
\begin{equation}
 \label{e:einstein_G}
G_{\alpha\beta} = 2T_{\alpha\beta} + 2 \left(
  \del_\alpha \psi \del_\beta \psi
  -\frac{1}{2}g_{\alpha\beta} g^{\mu\nu} \del_\mu \psi \del_\nu \psi \right)
\end{equation}
and 
\begin{eqnarray} \nonumber
\Tt_{\alpha\beta} ={}& \del_\alpha \phi\, \del_\beta \phi -
\frac{1}{2} \, \gt_{\alpha\beta} \, \gt^{\rho\sigma}
\del_\rho \phi\, \del_\sigma \phi, \\ \nonumber
T_{\alpha\beta} ={}& a^2(\psi) \Tt_{\alpha\beta} = a^2(\psi)
\big( \del_\alpha \phi \, \del_\beta \phi - \frac{1}{2}
g_{\alpha\beta} \, g^{\rho\sigma} \del_\rho \phi\,
\del_\sigma \phi \big).\nonumber
\end{eqnarray}

Following Christodoulou \cite{Christodoulou94}, we use the future-directed
null frame $(n, l)$, defined by
\begin{eqnarray}
&  n = 2e^{-\nu} \df{}{u} - e^{-\la} \df{}{r}, \nonumber
\\
&  l = e^{-\la} \df{}{r}. \nonumber
\end{eqnarray}
By contraction of the Einstein equations (\ref{e:einstein_G}) with $n,l$, we obtain the following system:
\begin{eqnarray}
\df{\la}{r} + \df{\nu}{r} 
 &=&
r \left( \df{\psi}{r} \right)^2  + r \left( a(\psi) \df{\phi}{r} \right)^2,
  \label{eq1} \\
\df{\la}{r} - \df{\nu}{r}
 &=&
\frac{1 - e^{2\la}}{r},
  \label{eq2} \\
 \qquad \df{\la}{u}
 &=&
r \left( \df{\psi}{r} \df{\psi}{u} 
       - e^{(\la - \nu)} \left( \df{\psi}{u} \right)^2 \right)
\nonumber \\
&&+ r a^2(\psi)  \left( \df{\phi}{r}
    \df{\phi}{u} - e^{(\la - \nu)} \left( \df{\phi}{u} \right)^2 \right).
  \label{eq3}
\end{eqnarray} 
Note that these equations involve only first-order derivatives of the metric and scalar fields.

On the other hand, by defining the function
\begin{equation}
  \label{e:def_coupling}
  \coupling(\psi) = {a'(\psi) \over a(\psi)},
\end{equation}
the evolution equation for the scalar field $\psi$ reads (cf.~again \cite{Damour92}) 
$$
\Box_g \psi = -\coupling(\psi) T 
=a'(\psi) a(\psi) \, g^{\alpha\beta}
\;\del_\alpha\phi\;\del_\beta\phi,
$$
where $\Box_g$ is the wave operator associated with the Einstein metric. 
In our gauge, this equation is equivalent to 
\begin{eqnarray}\nonumber 
  & -2 \left( \frac{\del^2 \psi}{\del r \del u} + \frac{1}{r}
    \df{\psi}{u} \right) + e^{(\nu - \la)} \left( \frac{\del^2
      \psi}{\del r^2} + \frac{2}{r} \df{\psi}{r} + \df{(\nu - \la)}{r}
    \df{\psi}{r} \right) \nonumber
  \\
  &  = a'(\psi) a(\psi) \left( e^{(\nu - \la)} \left(
      \df{\phi}{r} \right)^2 - 2 \df{\phi}{r}\df{\phi}{u} \right).\nonumber
\end{eqnarray}

Finally, the equation for the matter field $\phi$ is obtained from
the zero-divergence law for the stress-energy tensor:
$$
\Dt_\alpha \Tt^{\alpha\beta} = 0,
$$
where $\Dt$ is the covariant derivative associated with the physical metric.  
In terms of the Einstein metric, the zero-divergence law for the stress-energy tensor reads 
$$
\nabla_\alpha T^{\alpha\beta} = {a'(\psi) \over a(\psi)}  T \, \nabla^\beta \psi, 
$$
where $\nabla$ denotes the covariant derivative for the metric $g$. 
This equation is equivalent to
$$
\Box_g \phi = - {2 a'(\psi) \over a(\psi)} \, g^{\alpha\beta} \;\del_\alpha\phi\;\del_\beta\psi,
$$
a linear equation for $\phi$ which, in our gauge, becomes
\begin{eqnarray} \nonumber 
& -2 \left( \frac{\del^2 \phi}{\del r \del u} + \frac{1}{r}
    \df{\phi}{u} \right) + e^{(\nu- \la)} \left( \frac{\del^2
      \phi}{\del r^2} + \frac{2}{r} \df{\phi}{r} + \df{(\nu - \la)}{r}
    \df{\phi}{r} \right) \nonumber
\\
& =  -2\coupling(\psi) \left( e^{(\nu - \la)} \df{\psi}{r}
  \df{\phi}{r} -  \df{\phi}{r}\df{\psi}{u} - \df{\phi}{u}\df{\psi}{r}\right). \nonumber
\end{eqnarray}

%---------------------------------------------------------------------------------------------------

\subsection{Case of interest in this paper}

For definiteness and simplicity, we study the case in which
$\coupling(\psi)=\sigma$ is a {\sl constant}, which 
corresponds to Brans-Dicke
theory~\cite{Brans61}. Integrating (\ref{e:def_coupling}) we
get
$$
a(\psi) = a_0 \, e^{\coupling\psi},
$$
where $a_0$ is a dimensionless constant (independent of $\psi$).
Assuming that it does not vanish, this constant can be eliminated by re-defining the matter field $\phi$, 
and so we end up with 
$$
a(\psi) = e^{\coupling\psi},
$$
which is the choice made in the rest of this paper. 

Our equations can be easily compared with those used in other investigations of the 
Brans-Dicke theory. For instance, Liebling and Choptuik \cite{Liebling96}
have (denoting their variables with subindex LC)
$$
\phi = \psi_{LC}, 
\qquad 
\coupling\psi = -\frac{1}{2}\xi_{LC}, 
\qquad 
2\coupling^2 = (4\pi)\lambda_{LC},
$$
which imply $a^2 = e^{-\xi_{LC}}$. 
(Special care must be taken with the fact that, in \cite{Liebling96}, units
with $G=c=1$ are used so that various factors like $4\pi$ arise in their equations.) 

%===================================================================================================

\section{Self-similar assumption and the reduced system}
\label{section3}

\subsection{Essential field equations}

We now impose continuous self-similarity on the solutions, that is, 
\begin{equation}
  \label{homothetic}
  {\cal L}_S g_{\mu\nu} = 2 g_{\mu\nu}
\end{equation}
for some (conformal) homothetic Killing field denoted by $S^\mu$.  To work with
self-similar solutions, it is convenient to use adapted coordinates,
in which the integral lines of $S^\mu$ are now coordinate lines. 
Every spherically symmetric and self-similar spacetime (except Minkowski) has a singularity
at a point on the central world-line, and we shall use it to define the origin
of time, so that $u=0$ represents the future null cone of the singularity.
Moreover, since we are interested in the process of the formation of
singularities, we (principally) work in the past region $u<0$.

We define Bondi's self-similar coordinates by 
\begin{equation}
  \label{c44}
  x = \frac{r}{-u}, \qquad \tau = -\log(-u),
\end{equation}
where the sign in $\tau$ is chosen so that $u$ and $\tau$ increase
simultaneously toward the future. The homothetic vector is now $S=-\del_\tau$, with integral
lines $x=$const., and points away from the singularity ---which is now located
at $\tau=+\infty$.
Referring for instance \cite{GM03} for the general geometry of self-similar spacetimes, we note that 
in these coordinates the metric reads
\begin{equation}
\label{selfsimilarmetric}
g = e^{-2\tau} \left(
   (-e^{2\nu}+2x e^{\nu+\lambda})d\tau^2 
   -2e^{\nu+\lambda}d\tau dx
   + x^2 d\Omega^2
\right),
\end{equation}
where $\nu$ and $\lambda$ are functions of $(\tau,x)$. 
The Lie derivative ${\cal L}_S$ is now simply $-\partial/\partial\tau$ 
and, consequently, the symmetry condition (\ref{homothetic}) implies that all
metric coefficients depend on $x$, only, so
$$
\nu=\nu(x), \quad \qquad \lambda=\lambda(x).
$$
This implies similar conditions on the scalar fields 
and, since $\psi$ arises in an undifferentiated form, the
relevant condition is
$$
\frac{\partial}{\partial\tau} \psi = 0, 
\quad \mbox{ implying } \quad
\psi = \psi(x).
$$
However, $\phi$ only enters the equations in differentiated form and
hence the condition is
$$
\frac{\partial}{\partial\tau} \partial_\mu \phi = 0,
\quad \mbox{ implying } \quad
\phi = \chi(x) +  k \tau,
$$
where $\chi=\chi(x)$ is a function of $x$, only, and $k$ is a dimensionless real constant.

\

\noindent{\bf Remark.}
{\it 
An interesting variant of the above symmetry assumption 
was adopted in \cite{Hirschmann95},
where self-similar, complex-valued, scalar field solutions are constructed 
from the ansatz $e^{i\omega\tau}\xi(x)$, for some dimensionless real constant $\omega$.
}

\

Following Christodoulou~\cite{Christodoulou94}, we define 
\begin{eqnarray}
  \label{344} 
  \beta &= 1-\frac{e^{\nu-\lambda}}{2x}, 
  \nonumber\\
  \theta &= x\chi'(x), 
  \\
  \xi &= x\psi'(x),
  \nonumber
\end{eqnarray}
and we emphasize that $\beta$ will replace $\nu$ from now on. 
The Einstein equation (\ref{eq2}) becomes
$$
  x\frac{d\beta}{dx} = (1-\beta)(2-e^{2\lambda})
$$
and, by adding (\ref{eq1}) and (\ref{eq2}) together, we find
\begin{equation} 
  \label{lambdaeq}
  2x \, \frac{d\lambda}{dx} = \xi^2 + e^{2\coupling\psi}\theta^2-(e^{2\lambda} - 1).
\end{equation}
The third Einstein equation (in combination with the other two equations)
yields the constraint equation 
\begin{equation}
  \label{constraint}
  e^{2\lambda} = 1 + k^2\,e^{2\coupling\psi}
  + \frac{\beta}{1-\beta}\left((\theta+k)^2e^{2\coupling\psi}+\xi^2\right).
\end{equation}
Moreover, the wave equation for the matter field reads 
\begin{equation} 
  \label{dthetadx}
  \beta x \frac{d\theta}{dx} + \left( 1 - (1-\beta)e^{2\lambda}\right) \, \theta
  + k = -\coupling(k+2\beta\theta)\xi,
\end{equation} 
and the wave equation for the Brans-Dicke field is
\begin{equation} 
  \label{dxidx}
  \beta x \frac{d\xi}{dx} + \big( 1 - (1-\beta)e^{2\lambda}\big) \, \xi 
  =
  \coupling e^{2\coupling\psi}(k+\beta\theta)\theta.
\end{equation}
(We emphasize that there is {\sl no} factor $2$ in the parenthesis of the
right-hand side of this last equation.)

In conclusion, using the constraint (\ref{constraint}) to eliminate $\lambda$,
we arrive at a system of four ordinary differential equations: 
\begin{eqnarray}
  \label{mainsyst} 
  x\frac{d\beta}{dx} 
  &=
  1-k^2e^{2\coupling\psi} - \big( e^{2\coupling\psi}(\theta+2k)\theta
  +\xi^2+1\big) \, \beta,  
  \\
  \beta x\frac{d\theta}{dx} 
  &=
  k(ke^{2\coupling\psi}\theta-1)
  -\coupling(k+2\beta\theta)\xi
  +\big( e^{2\coupling\psi}(\theta+2k)\theta +\xi^2-1\big) \, \beta\theta, 
  \nonumber \\
  \beta x\frac{d\xi}{dx} 
  &=
  k^2 e^{2\coupling\psi}\xi
  +\coupling e^{2\coupling\psi}(k+\beta\theta)\theta
  +\big( e^{2\coupling\psi}(\theta+2k)\theta +\xi^2-1 \big) \, \beta\xi, 
  \nonumber \\
  x\frac{d\psi}{dx} &= \xi.\nonumber
\end{eqnarray}
The first two equations above reduce to Christodoulou's equations (see
(0.27a) and (0.27b) therein) {\sl provided}
$\coupling$ and $\xi$ vanish identically. 
Note that we can simultaneously change the signs of $\theta$ and
$k$ without changing the structure of the system. Hence, without 
loss of generality,  we can assume that $k\ge 0$, while $\theta$ still can have any sign. 

%---------------------------------------------------------------------------------------------------

\subsection{Reduced system}  

We have found it convenient to rescale $\theta$ with an exponential factor
$e^{\coupling\psi}$, which compensates for the discrepancy by a factor $2$
between equations (\ref{dthetadx}) and (\ref{dxidx}) (as pointed out earlier).
We define
$$
  \Theta = e^{\coupling\psi}\theta, 
$$
and the remaining exponential terms can be combined with $k$ into a single variable: 
$$
  K = e^{\coupling\psi} k.
$$
The notation is intended to compare with Christodoulou's case, for which
$\Theta$ and $K$ coincide with $\theta$ and $k$, respectively.
Using the variable $s=\ln x$ and denoting $d/ds$ as a prime,
the system (\ref{mainsyst}) now reads
\begin{eqnarray}
  \label{reducedbeta}
  \beta' &=
  1-K^2 -\big(2K\Theta+\Theta^2+\xi^2+1\big) \, \beta,
  \nonumber\\
  \beta\,\Theta' &=
  K\big(K\Theta-1\big) - \coupling\big(K+\beta\Theta\big)\, \xi
  + \big(2K\Theta+\Theta^2+\xi^2-1\big)\,\beta\Theta,
  \\
  \beta\,\xi' &=
  K^2\xi + \coupling\big(K+\beta\Theta\big)\, \Theta
  + \big(2K\Theta+\Theta^2+\xi^2-1\big)\,\beta\xi,
  \nonumber\\
  K' &= \coupling K \xi. 
  \nonumber
\end{eqnarray}

It is also convenient to make the change of variable 
$$
  \alpha=\frac{1}{\beta},
$$
such that the system under consideration becomes polynomial in all variables:
\begin{eqnarray}
  \label{reduced}
  \alpha' &=
  \big( \alpha K^2+2K\Theta+\Theta^2+\xi^2+1\big) \, \alpha - \alpha^2, 
  \nonumber\\
  \Theta' &=
  \big( \alpha K^2+2K\Theta+\Theta^2+\xi^2-1\big) \, \Theta
  -\coupling (\alpha K+\Theta)\xi -\alpha K, 
  \\
  \xi' &=
  \big(\alpha K^2+2K\Theta+\Theta^2+\xi^2-1\big) \, \xi
  +\coupling (\alpha K+\Theta)\Theta, 
  \nonumber\\
  K' &= \coupling K \xi. 
  \nonumber
\end{eqnarray}
From now on, we refer to these equations as 
the {\sl reduced system}, which is our main object of study.
We sometimes use it in the form (\ref{reducedbeta}),
evolving $\beta$ instead of $\alpha$.

Observe the combination
\begin{equation}
  \label{e2lambda}
  \alpha K^2 +2K \Theta + \Theta^2 + \xi^2 =
  (\alpha-1)(e^{2\lambda}-1).
\end{equation}
We will later use the variable
\begin{displaymath}
L = \sqrt{\Theta^2+\xi^2},
\end{displaymath}
that obeys the evolution equation
\begin{equation}
  \label{eqL}
  \frac{1}{2} (L^2)' 
  =
  \big( \alpha K^2+2K\Theta+ L^2 -1\big) \, L^2 - \alpha K \Theta.
\end{equation}

\

\noindent{\bf Remark.} {\it 
  From $\Theta$ and $\xi$ one can form a complex function 
$\Lambda = \Theta + i\xi\label{eqLambda}$,
  with norm $|\Lambda|=L$, which satisfies the differential
  equation
$$ 
    \Lambda' = \alpha \left( Z_1 - Z_0 \right) \left( Z_0
      \Lambda - 1 \right) + \left| \Lambda \right| \left( \Lambda +
      \frac{Z_1}{2} \right) + \frac{\Lambda^2 Z_1}{2} -
    \Lambda  
$$
  with $Z_0 = K + i\coupling$ and $Z_1 = 2K + i\coupling$. 
}

%===================================================================================================

\section{Interior solution originating at the center}
\label{section4}

\subsection{Analytical strategy}

The construction of our spacetimes, as solutions of the reduced system
(\ref{reduced}), will be performed in several steps; a main difficulty stems from the fact that
 the equations become
singular for several values of the $x$-coordinate. This happens at the center of spherical symmetry, $x=0$, and at the
{\sl self-similarity horizons} corresponding to those values of $x$ for which the
homothetic vector $S$ becomes null. In our Bondi
coordinates, this corresponds to the condition
$g(\partial_\tau,\partial_\tau)=g_{\tau\tau} = 0$ for the past light-cone,
which reads (see (\ref{selfsimilarmetric}))
$$
-e^{2\nu} + 2 x e^{\nu+\lambda} = 0, 
\qquad \textrm{or} \quad \beta = 0.
$$
This section describes the construction of the past of the singularity,
namely the region between the center worldline and the past lightcone of
the singularity, the first self-similarity horizon. Following Christodoulou,
we shall refer to this region as {\sl the interior solution}.

\begin{figure}[ht!]
\psfrag{Interior}{Interior}
\psfrag{Exterior}{Exterior}
\psfrag{Future}{Future}
\psfrag{x=0}{$\!\!\!\!\!\!x=0$}
\psfrag{x=x*}{$x=x_*$}
\psfrag{x=inf}{$\!\!x=+\infty$}
\begin{center}
\includegraphics[width=5cm]{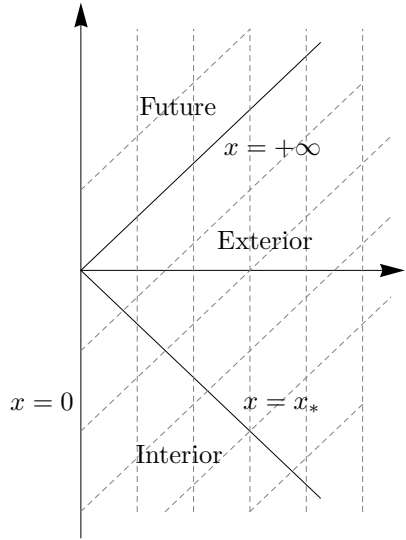}
\end{center}
\caption{\label{regions}
Schematic representation of the different regions in which the
the solution is divided, matching at the past light-cone ($x=x_*$) and
the future light-cone ($x=+\infty$). The dashed lines denote Bondi coordinate
lines.
}
\end{figure}

%--------------------------------------------------------------------------------------------------
 
\subsection{Critical points and regularity at the center}\label{ss:center}

\begin{figure}
\psfrag{theta}{$\Theta$}
\psfrag{xi}{$\xi$}
\begin{center}
\includegraphics[width=8cm]{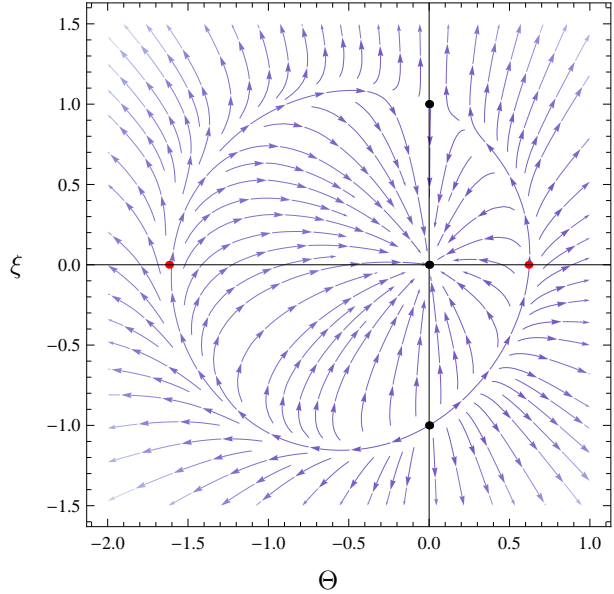}
\end{center}
\caption{\label{streams}
Projection onto the plane ($\alpha=0$, $\Theta$, $\xi$, $K=1/2$) of 
streamlines of the flow vector field of system (\ref{reduced}) for
$\coupling = 1/3$.
We have marked the fixed point (a) at $\Theta=0,\xi=0$ and the projections
at $\Theta=0,\xi=\pm 1$ of the exact solutions mentioned in the text.
We have also marked the points $\Theta = -K\pm \sqrt{K^2+1}$, which are
fixed points in Christodoulou's system, but are no longer fixed points
in our system. Those four points now form part of an unstable 
projected structure that resembles a limit circle, though we do not
know whether a true limit cycle exists in the full phase space.
}
\end{figure}

We begin by determining all critical points corresponding to equilibria of the
reduced system, provided $\sigma\not=0$. In view of the right-hand side of
(\ref{reduced}) and provided the unknown functions $\alpha, \Theta, \xi, K$
have vanishing derivatives, only the following alternatives can arise:  
\begin{itemize}

\item[(a)] $\alpha=0$, $\Theta=0$, $\xi=0$, $K$ arbitrary,

\item[(b)] $\alpha=1$,  $\Theta=-K$, $\xi=0$, $K$ arbitrary,

\item[(c)] $\alpha=0$, $\Theta=0$, $\xi^2=1$, $K=0$,

\item[(d)] $\alpha=2$, $\Theta=0$, $\xi^2=1$, $K=0$.
\end{itemize}
Points (c) actually belong to the exact solutions $\alpha=0, \Theta=0,
\xi=\pm 1, K= K_0 e^{\pm \coupling s}$. Note that this collection of
fixed points is not an extension of Christodoulou's result. This is
due to the fact that the condition $\xi=0$ is not preserved by the evolution.
In fact, several fixed points in Christodoulou's problem are no longer
fixed in our case. This is clear in Figure~\ref{streams}, which shows
a projection of the evolution flow on a slice $(\Theta,\xi)$ of phase space.

The center of symmetry must correspond to one of the above cases. We need
to rely on physically motivated regularity requirements, to select one of
them. Specifically, we impose that the center of symmetry is regular before the formation
of the singularity. Recall that the Hawking mass is determined from
the metric coefficient $\lambda$ by the relation 
\begin{equation}
  1-\frac{2m}{r} = e^{-2\lambda}. 
\end{equation}
In order to avoid a singular behavior at the center, we impose
that the mass tends to zero, which implies $\lambda=0$ at the center.
The equation (\ref{e2lambda}) then selects the fixed points (a) and (b), above.
 
Regularity at the center also requires that $\nu(x=0)$ is finite, to avoid a
coordinate time singularity.  However, the value of $\nu$ at the center is
gauge-dependent and, by normalizing $\nu$ to vanish at the center, the
coordinate time $u$ in (\ref{c44}) coincides with the proper time of the
central observer.   Consequently, $\beta$ in (\ref{344}) behaves like $\beta
\sim -1/(2x)$ and, equivalently $\alpha \sim -2x$ at the center.  In
particular, this condition implies that $\alpha=0$ at the center, which is
consistent with the critical point (a) above, only.

In summary, we obtain the following values for the critical point of interest: 
\begin{equation}
  \label{399}
  \alpha=0, \quad \Theta=0, \quad K=\kk, \quad \xi=0 \qquad \textrm{ at the center } x=0,
\end{equation}
where $\kk$ is an arbitrary non-negative constant.

%---------------------------------------------------------------------------------------------------

\subsection{Linear stability of the critical point at the center} 
\label{ss:exists}

After linearizing around the critical point (a), 
the Jacobian matrix of the system in the linearized variables $(\delta\alpha,\delta\Theta,\delta\xi,\delta K)$
reads 
$$
  \left(\begin{array}{cccc}
      1    & 0  & 0  & 0 \\
      -\kk & -1 & 0  & 0 \\
      0    & 0  & -1 & 0 \\
      0    & 0  & \sigma \kk  & 0
    \end{array}\right). 
$$ 
Its
eigenvalues are $1, -1, -1, 0$, with respective eigenvectors
$$
%\hspace{-2.5cm}  
  e^{(1)} = \left(\begin{array}{c} 1 \\ -\kk/2 \\ 0 \\ 0 \end{array}\right), \quad
  e^{(2)} = \left(\begin{array}{c} 0 \\ 1 \\ 0 \\ 0 \end{array}\right), \quad
$$
$$
  e^{(3)} = \left(\begin{array}{c} 0 \\ 0 \\ 1 \\ - \sigma k \end{array}\right), \quad
  e^{(4)} = \left(\begin{array}{c} 0 \\ 0 \\ 0 \\ 1 \end{array}\right), 
$$
respectively. 
(Figure~\ref{streams} corresponds to the two negative eigenvalues.)
The origin (a), therefore, has an unstable branch, tangent to the vector $e^{(1)}$, and 
the corresponding solutions in the neighborhood of (a) have the form
$$
  \left(\begin{array}{c} \alpha \\ \Theta \\ \xi \\ K \end{array}\right)
  = \left(\begin{array}{c} 0 \\ 0 \\ 0 \\ \kk \end{array}\right)
  + a_1\, e^s \left(\begin{array}{c} 1 \\ -\kk/2 \\ 0 \\ 0 \end{array}\right) + o(e^s),
$$
where $a_1$ is a parameter. 

Imposing the normalization $\alpha\, x^{-1} \to -2$ at (a), we get $a_1 = -2$
and we obtain an interior solution in the neighborhood of the center, 
satisfying 
\begin{equation}
  \left(\begin{array}{c} \alpha \\ \Theta \\ \xi \\ K \end{array}\right)
  = \left(\begin{array}{c} - 2 e^s \\ \kk e^s \\ 0 \\ \kk  \end{array}\right)
  + o( e^s).\label{e:Taylor_0}
\end{equation}

We now show that this implies that the spacetime is (mildly) singular at the center.
We start by
taking the trace of (\ref{e:einstein_G}), which gives the Ricci scalar
\begin{equation}
  \label{e:Ricci}
  R = -2T + 2\left( g^{\alpha\beta} \partial_\alpha \psi \partial_\beta \psi \right). 
\end{equation}
Taking the form~(\ref{selfsimilarmetric}) of the metric, we get
\begin{eqnarray}
  R &= &2e^{2\tau} \left( -2e^{-(\nu + \la)}
     \left( a^2\partial_\tau \phi \partial_x
      \phi+ \partial_\tau \psi \partial_x \psi \right)  \right.\\
   & &\left. + \left( e^{-2\la} - 2x e^{-(\nu + \la)} \right)
     \left( a^2\left( \partial_x \phi \right)^2
      + \left( \partial_x \psi \right)^2 \right) \right)
\nonumber
\\
 &=& -\frac{2 e^{2\tau} \left(\Theta
   K+\beta \left(\Theta^2+\xi^2\right)\right)}
   {x^2 \left(2 \beta\Theta K+K^2
     +\beta \left(\Theta^2+\xi^2-1\right)+1\right)}. 
\label{RicciscalarB}
\end{eqnarray}
Using the reduced system~(\ref{reduced}) together with the 
expansion~(\ref{e:Taylor_0}), we get (for the first two variables) 
\begin{eqnarray*}
  \alpha &={} -2\,e^s -4e^{2s} + o\left(e^{2s}\right),\\
  \Theta &={} K_0\, e^s + \frac{4K_0}{3}\, e^{2s} + o\left(e^{2s}\right).
\end{eqnarray*}
Therefore, the local expression for the Ricci scalar becomes (we have replaced
 $e^s$ with $x$) 
\begin{equation}
  \label{e:R_origin}
  R = e^{2\tau} \left( -2K_0^2 - 4K_0^2x + o \left( x^2 \right) \right),
\end{equation}
where we have used 
$
\lim_{x\to 0} \nu(x) = \lim_{x\to 0} \la(x) = 0.
$

Consequently, the presence of a non-vanishing linear term $-4K_0^2x$ in (\ref{e:R_origin})
shows that, when viewed as a geometric object in a spherically symmetric spacetime,
{\sl the scalar curvature $R$ is continuous but not differentiable} 
at the center $x=0$. (Only even powers of $x$ should, otherwise, be allowed.) 
Hence, the spacetime contains a (mild) singularity
before the central curvature singularity forms at $u=0$.

%---------------------------------------------------------------------------------------------------
 
\subsection{Integration in the interior region}

We denote by $(-\infty, s_*)$ the maximal interval on which the solution is
defined, and 
now check that, provided $s_*$ is finite, $\alpha$ must blow-up.

\

\noindent{\bf Claim 1.} {\it  
  Either the solution is defined and regular up to $s_*=+\infty$, or else
  $\alpha(s) \to -\infty$ but $\Theta, \xi$ and $K$ remain bounded as $s
  \to s_*$.
}

\ 

Indeed, se suppose that $s_*$ is finite and we are going to
prove that both $\Theta$ and $\xi$ are bounded. First, $\alpha <0$ for all $s < s_*$. Indeed, $\alpha(s) <0$ at least for an
interval of the form $(-\infty, s_0]$ with $s_0 < s_*$. Thus, $\beta(s) =
1/\alpha(s)$ is negative and finite on the same interval.  Also, setting $B= 2
K \Theta + \Theta^2 + \xi^2 + 1$ and $C = 1 - K^2$, from the first equation
of (\ref{reducedbeta}) we deduce 
that (for $ s_0\leq s < s_*$) 
\begin{displaymath} 
\beta (s) = \beta(s_0) e^{-\int_{s_0}^s B(u) du} + \int_{s_0}^s C(u)
e^{-\int_{u}^{s} B(t) dt} du.
\end{displaymath}
Thus, $\beta(s)$ remains finite on $(-\infty, s_*)$ so that $\alpha(s)=
1/\beta(s)$ can not vanish in this interval and remains negative.
Now, the equation (\ref{lambdaeq}) reads 
\begin{eqnarray*}
2 \lambda' & = (\xi^2 + \Theta^2\big) - (e^{2\lambda}-1)
\\
& = L^2 - (e^{2\lambda}-1),
\end{eqnarray*}
thus we have 
$$ 
  e^{2\lambda} = {{e^s g(s)} \over {\int_{-\infty}^s e^{s'} g(s') \, ds'}},           
  \qquad
  g(s) = e^{\int_{-\infty}^s L^2(s') \, ds'}.  
$$

But, $L$ can not vanish, except at an isolated point. Indeed, if there is 
$s_1 < s_*$ such that $L(s_1) = 0$, then $\Theta(s_1)=0$ with 
$\Theta'(s_1) = - \alpha(s_1) K(s_1) >0$.
Thus, functions $\Theta$ and $L$ are both different from zero in the
neighborhood of $s_1$. We obtain that the function $g$ is strictly monotone
increasing,
\begin{displaymath}
\int_{-\infty}^s e^{s'} g(s') \, ds' < g(s) \int_{-\infty}^s e^{s'} \, ds' =
e^s g(s),
\end{displaymath}
so that $e^{2\lambda} > 1$, i.e., $\lambda(s) >0$ for all $s < s_*$.
Setting 
\begin{displaymath} 
\Gamma = \alpha K^2 + 2\Theta K + L^2,
\end{displaymath}
and using (\ref{constraint}), we obtain 
\begin{displaymath} 
e^{2\lambda} - 1 = {\Gamma\over {\alpha - 1}}.
\end{displaymath}
Since $\alpha<0$ for $s<s_*$, we get $\Gamma(s) < 0$ for all
$s\in (-\infty, s_*)$.

Let us introduce now the quantity
\begin{displaymath} 
H = \Theta^2 + \xi^2 + 2{\xi \over \sigma} =  L^2 + 2{\xi \over \sigma}.
\end{displaymath}
Using (\ref{eqL}) and the third equation in (\ref{reduced}), we obtain
\begin{displaymath} 
H' = \Gamma H + \Gamma L^2 - 2 ( \xi^2 + {\xi\over \sigma}).
\end{displaymath}
Then, completing the square in the last term,
\begin{displaymath} 
H' \leq \Gamma H + {1\over 2 \sigma^2}.
\end{displaymath}
Now, fix $s_0 < s_*$. Using the Gronwall inequality, we get for $s \in [s_0, s_*)$,
\begin{displaymath} 
H (s) \leq H(s_0) e^{\int_{s_0}^s \Gamma (u) du} + { 1\over
  {2\sigma^2}}\int_{s_0}^s e^{\int_{u}^{s} \Gamma(t) dt} du\end{displaymath}
\begin{displaymath}
\leq H^+(s_0) + { 1\over {2\sigma^2}}( s - s_0) \leq H^+(s_0) + { 1\over
  {2\sigma^2}}( s_* - s_0),
\end{displaymath}
where $H^+(s_0) = {\rm max}(0, H(s_0))$.  We deduce that both $\Theta$ and $\xi$ are
bounded, and since $K(s) = K(s_0)e^{\int_{s_0}^s \xi}$, $K$ is also bounded.
We conclude that necessarily $\alpha\to -\infty$, as $s\to s_*$.

\

\noindent{\bf Claim 2.} {\it  
  Assume that $s_* < +\infty$. Then, there exists a real $0 < K_* < 1$ such
  that,
  \begin{eqnarray*}
    \lim_{s\to s_*} K(s) &={} K_*,\\
    \lim_{s\to s_*} \Theta(s) &={} \frac{K_*}{K_*^2 + \coupling^2},\\
    \lim_{s\to s_*} \xi(s) &={} -\frac{\coupling}{K_*^2 + \coupling^2}.
  \end{eqnarray*}
}

To establish this claim, we proceed as follows. 
First, since $\xi$ remains bounded, $\int_{-\infty}^{s_*} \xi(s) ds = \Xi$,
where $\Xi$ is some real constant, so that
\begin{displaymath}
\lim_{s\to s_*} K(s) = \kk e^{\sigma\,\Xi} = K_*,
\end{displaymath}
with $K_* > 0$. Also, the first equation of (\ref{reducedbeta}) gives
$
\beta'(s) \to 1 - K_*^2,
$
and we obtain
\begin{equation} 
  \label{behavbeta} 
  \frac{\beta(s)}{s_* - s}\to K_*^2 - 1.
\end{equation}
But since $\beta(s) < 0$ for $s < s_*$, we get that necessarily $K_* \leq
1$. The case $K_* = 1$ will be excluded at the end of the argument.

Now, let us introduce the new variables  
\begin{equation}
  \label{delta12}
  \delta_1 = K\Theta - \sigma \xi - 1,\,\,\,\,\,\, \delta_2 = K \xi + \sigma \Theta.
\end{equation}
Thus, the expressions of $\Theta'$ and $\xi'$ in (\ref{reduced}) become
\begin{equation}
  \label{thetaxidelta}
  \Theta' ={K\over \beta} \delta_1  + h_1,\,\,\,\,\,
  \xi' = {K\over \beta} \delta_2 + h_2,
\end{equation}
where
  \begin{eqnarray} \nonumber
& h_1  = (2 K \Theta + \Theta^2 + \xi^2 - 1)\Theta -\sigma \xi\Theta,\ \nonumber
\\ \nonumber
& h_2  = (2 K \Theta + \Theta^2 + \xi^2 - 1)\xi + \sigma\Theta^2. \nonumber
  \end{eqnarray}
Using (\ref{delta12}), (\ref{thetaxidelta}) and the last equation in
(\ref{reduced}), we get
\begin{equation}
  \label{delta12prime}
  \delta_1' = {K\over \beta}(K \delta_1 - \sigma \delta_2) + l_1,\,\,\,\,
  \delta_2' = {K\over \beta}(\sigma \delta_1 + K \delta_2) + l_2,
\end{equation}
where
  \begin{eqnarray}
  \label{l12}
  l_1 &= \sigma K \xi \Theta + K h_1 - \sigma h_2,
  \\
  l_2 &= \sigma K \xi^2 + K h_2 + \sigma h_1.
  \end{eqnarray}
Thanks to (\ref{delta12prime}), the quantity $\delta = (\delta_1^2 +
\delta_2^2)^{1/2}$ satisfies
\begin{equation}
  \label{deltaprime}
  \delta' = {K^2\over \beta}\delta + l,
\end{equation}
where
$$
l= \frac{l_1 \delta_1 + l_2 \delta_2}{ \delta}.
$$

By assumption, $l_1$ and $l_2$ are bounded on $(-\infty, s_*)$ so that $l$ is
bounded too.  Choosing $s < s_*$, and integrating (\ref{deltaprime}) over
$[s_0, s]$ we obtain
$$
\delta(s) = e^{-\zeta(s)}\Big( \delta(s_0) + \int_{s_0}^s e^{\zeta(s')} \, l(s') ds'\Big),
$$
where
$$
 \zeta(s) = - \int_{s_0}^s \frac{K^2(s')}{ \beta(s')} ds'. 
$$
Function $\zeta$ is increasing, and by (\ref{behavbeta}), it tends to $+\infty$
as $s \to s_*$, together with its derivative $\zeta'$. Thus,
$$
e^{-\zeta(s)}\int_{s_0}^s e^{\zeta(s')} ds'\to 0 \, \, as \,\, s \to s_*,
$$
and since $l$ is bounded we obtain 
$$
\lim_{s\to s_*} \delta(s) = 0.
$$
Finally, we can rewrite (\ref{delta12}) in the form
$$
\Theta = {1\over \sigma^2 + K^2} (K(\delta_1 + 1) + \sigma \delta_2),
\qquad 
\qquad 
\xi = {1\over \sigma^2 + K^2} (-\sigma(\delta_1 + 1) + K \delta_2),
$$
and we get 
$$
\lim_{s\to s_*} \Theta(s) = \frac{K_*}{K_*^2 + \coupling^2},
\qquad \qquad 
\lim_{s\to s_*} \xi(s) = - \frac{\coupling}{K_*^2 + \coupling^2}.
$$

Now, it remains to prove that $K_* < 1$, i.e., $K_* = 1$ is excluded when $s_*
< +\infty$, as well as $K_* >1$.  So, assume by contradiction that $K_* =
1$. Thus, the last equation in (\ref{reduced}) reads
\begin{equation}
\label{Kequals1}
K(s)=  1 - \frac{\sigma^2}{\sigma^2 + 1} ( s- s_*)  + o(|s - s_*|), 
\end{equation}
Also, (\ref{reducedbeta}) gives $\beta'(s) \to 0$ as $s\to s_*$, so that $\beta(s) = o (|s_* - s|)$.
More precisely, using (\ref{Kequals1}) in the first equation of (\ref{reducedbeta})  we obtain the following expansion
$$
\beta'(s) =   2 \frac{\sigma^2}{\sigma^2 + 1} ( s - s_*)  + o(|s - s_*|).
$$
Thus, $\beta$ is decreasing in the neighborhood of $s_*$ which is impossible
since $\beta <0$ and $\beta(s) \to 0$ as $s\to s_*$.

%--------------------------------------------------------------------------------------------------

\subsection{Conclusions for the interior region} 

In the interior region, the situation is similar to that of Christodoulou, in the sense that
the presence of a first self-similarity horizon, the past light-cone of the
singularity, is determined by the value of a constant $K_*$, which must be
{\sl below the threshold $1$.} The main difference is that in Christodoulou's case this
constant is the parameter $k$, fixed throughout the problem, while
 in our
case the constant $K_*$ is dynamically determined by the evolution
and, so, depends on the set of initial conditions, especially the initial value $\kk$ of the
variable $K$.

\begin{figure}
  \psfrag{coordx}{$x$}
  \psfrag{KK}{$K$}
  \centerline{\includegraphics[width = 10cm]{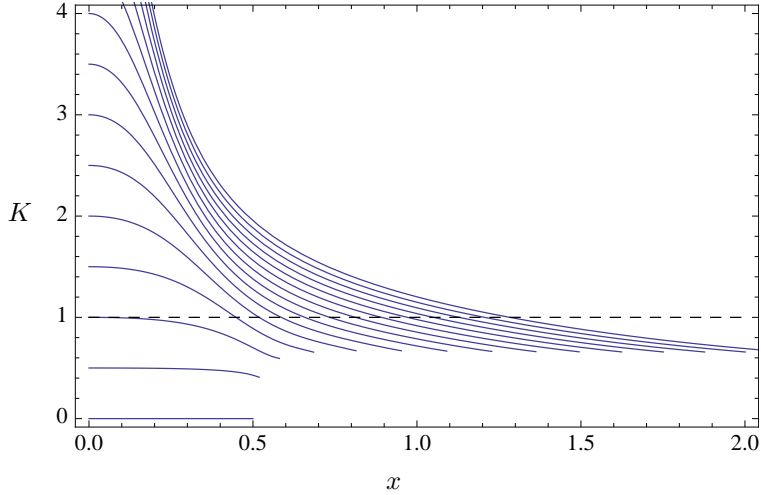}}
  \caption{\label{Kinterior}
    Evolution of $K(x)$ starting from values $K_0$ ranging from $0$ to $7$
    in steps $1/2$, hence $15$ curves. The coupling constant is $\sigma=3/2$
    in all cases. Each evolution stops when reaching the singular point $x_*$,
    which for $K_0=0$ corresponds to Minkowski, with $x_*=1/2$.
    For reference, a dashed line represents the value $K=1$.}
\end{figure}

We have performed numerical integrations of the
reduced system of equations to investigate how $K$ evolves. The results
are summarized in Figure~\ref{Kinterior}, which shows several evolutions
of $K$ starting from different values at the center. In all cases we see
a decrease of $K$ until {\sl values below $1$} are reached, and then we reach the
singular point $x_*=e^{s_*}$, where the integration is stopped. We have found
this behavior in all tested cases, including cases with large values of the initial $K_0$ (above 1000, say). 
The value of $\sigma$ does not alter the qualitative picture,
though the decay of $K$ is faster for larger values of $\sigma$. 
Interestingly enough, 
for large values of $K_0$ the final value $K_*$ is almost independent of that initial
value.  For sufficiently large values of $\sigma$ we find numerically that $K_*$ tends to
$4/(3\sigma)$.

In other words, we can have solutions in which the central singularity has a
past light-cone for initial values of the constant $k$ for which Christodoulou's
corresponding solution would be more pathological, with that light-cone becoming
actually a border of the spacetime.

%===================================================================================================

\section{Extension to the exterior region} 
\label{section5}

\subsection{The singular points}

According to the previous section when $s_* < +\infty$, the solution of
(\ref{reduced}) converges to a singular point of the form $( \beta, \Theta,
\xi, K) = (0, \Theta_*, \xi_*, K_*) = U_{K_*}$, where $ 0 < K_* < 1$,
$\Theta_* = \frac{K_*}{K_*^2 + \coupling^2}$, $\xi_* = - \frac{\sigma}{K_*^2
  + \coupling^2}$.  To treat the solutions in the neighborhood of such
singular point, and following \cite{Christodoulou94}, we introduce a new
independent variable t satisfying
\begin{displaymath} 
{ds\over dt} = \beta,
\end{displaymath}
which converts the singular point at finite $s$ into a critical point at $t\to -\infty$.

Thus, by using (\ref{reducedbeta}), the variables $\beta, \Theta, \xi,$ and $K$
satisfy the system
\begin{eqnarray}
  \label{singularsystem}
  {d \beta\over d t}&= (1 - K^2) \beta - ( 2 K \Theta + \Theta^2 + \xi^2 + 1)\beta^2, 
  \nonumber\\
  {d \Theta\over d t} &= \beta\, \Theta (  2 K \Theta + \Theta^2 + \xi^2 - 1 -
  \sigma \xi) + K^2 \Theta - K - \sigma K \xi, 
  \\
  {d \xi\over d t} &= \beta\, \xi\, (  2 K \Theta + \Theta^2 + \xi^2 - 1) +
  \sigma\, \beta \,\Theta^2 + K^2\xi + \sigma K \Theta, 
  \nonumber\\
  {dK\over d t} &= \coupling K \beta\, \xi.\nonumber
\end{eqnarray}
The singular point $U_{K_*}$ is an equilibrium point of the previous system,
and the Jacobian matrix at this point reads
\begin{displaymath}
A( U_{K_*})= \left( 
  \begin{array}{cccc}
    1 - K_*^2 & 0 & 0 & 0 \\
    \\
    {K_*^2( K_*^2 + 1)\over (K_*^2 + \sigma^2)^2}  & K_*^2 &  - \sigma K_* & 
    {K_*^2\over K_*^2 + \sigma^2}\\
    \\
    {\sigma( 1 - \sigma^2)\over (K_*^2 + \sigma^2)^2}& \sigma K_* & K_*^2 &
    -{\sigma K_*\over K_*^2 + \sigma^2}\\ 
    \\
    -{\sigma^2 K_*\over K_*^2 + \sigma^2}& 0 & 0 & 0 
  \end{array}
\right). 
\end{displaymath} 

The spectrum  of $A( U_{K_*})$ is  given by
\begin{displaymath} 
Sp(A( U_{K_*})) = \{0,\,\, 1 - K_*^2,\,\, K_*^2 - i \sigma K_*,\,\, K_*^2 + i
\sigma K_*\}.
\end{displaymath}
The eigenvalue $0$ corresponds to the fact that the set ${\cal C}_* =\{
U_{K_*}, 0 < K_* <1 \}$ defined by the equilibrium points of the form
$U_{K_*}$, with $ 0 < K_* <1$, is a (one-dimensional) curve. Each point
$U_{K_*}$ has an unstable manifold $W_i^{K_*}$ of dimension three,
corresponding to the three other eigenvalues having a positive real part.
Naturally, $W_i^{K_*}$ must be transverse to ${\cal C}_*$. 

We also observe that all solutions originating at $U_{K_*}$ and 
extending from the interior to the exterior region,
admit the following expansion (when $t\to -\infty$): 
\begin{eqnarray} 
\nonumber
  \left(\begin{array}{c} \beta \\ \Theta \\ \xi \\ K \end{array}\right)
  =  
& U_{K_*}  + a_1 e^{(1 - K_*^2)t} C_1  + a_2 e^{ K_*^2t} C_2(t)  + a_3 e^{ K_*^2t} C_3(t) \nonumber
\\\nonumber
& + o(e^{ K_*^2t},e^{(1 - K_*^2)t} ),\nonumber
\end{eqnarray}
where $C_2(t)$ and $C_3(t)$ are (bounded) periodic-rotating vector-valued functions, 
and $C_1$ a fixed eigenvector of the matrix $A(U_{K_*})$ corresponding
to the eigenvalue $\lambda_1 = 1 - K_*^2$. The three vectors $C_1$, $C_2(t)$ and $C_3(t)$ are linearly independent for all $t$.
Up to a translation in $t$, we can assume that $a_1 = 1$ and, thus, we obtain a two-parameter family of solutions.

The expansion above shows that the functions on the left-hand-side are
{\sl continuous at the past light-cone, but not differentiable in $s$} 
there for $0<K_*<1$, since they will generically contain terms of the form
$(s-s_*)^{K_*^2/(1-K_*^2)}$ or $(s-s_*)^{1-1/K_*^2}$. In particular $\Theta$
will be only continuous, and hence {\sl curvature will be discontinuous,} though
still finite, at the past light-cone.

%---------------------------------------------------------------------------------------------------

\subsection{Exterior solutions}

We computed in Section \ref{ss:center} the fixed points of our dynamical
system. In the interior the relevant point was (a), but now we will study
(b) and (d).
We first focus our study of the exterior region on an equilibrium point of
the form (b), namely 
$$
P_{K^*} = (1, -K^*, 0, K^*) \qquad \mbox{ for a fixed } 0 < K^* < 1.
$$ 
The Jacobian matrix of system (\ref{reduced}) at this point is given by
\begin{displaymath}
A(P_{K^*})= \left(
  \begin{array}{cccc}
    {K^*}^2 - 1& 0 & 0 & 0 \\
    - {K^*}^3 - {K^*} & - 1 & 0 & - 1
    \\
    - \sigma {K^*}^2 & - \sigma {K^*} & - 1& - \sigma {K^*} \\
    0 & 0 & \sigma {K^*} & 0
  \end{array}
\right). 
\end{displaymath} 
The spectrum of this matrix is given by
\begin{displaymath}
Sp(A(P_{K^*})) = \{0,\,\, {K^*}^2 - 1,\,\, - 1 - i \sigma {K^*},\,\, - 1 + i \sigma {K^*} \}, 
\end{displaymath}
which gives the asymptotic behavior
\begin{eqnarray}
\hspace{-2cm}
\label{expansionFLC}
  \left(\begin{array}{c} \alpha \\ \Theta \\ \xi \\ K \end{array}\right)
  =  
& P_{K^*}  + a_1 e^{({K^*}^2-1)s} C_1  + a_2 e^{ (-1-i\sigma K^*)s} C_2(t)  
\nonumber
\\
& + a_3 e^{ (-1+i\sigma K^*)s} C_3(s) + o(e^{-s},e^{({K^*}^2-1)s} ),
\end{eqnarray}

The eigenvalue $0$ corresponds to the fact that the set ${\cal C}^* =
\{ P_{K^*}, 0 < K^* <1 \}$ defined by the equilibrium points of the
form $P_{K^*}$, is a (one-dimensional) curve.
Each point $P_{K^*}$ has a stable manifold $W_s^{K^*}$ of dimension three,
corresponding to the three other eigenvalues having
a negative real part. Naturally, $W_s^{K^*}$ must be transverse to ${\cal C}^*$. 

Consider now the two isolated critical points of form (d)
$$
V_\epsilon = (2, 0, \epsilon, 0),\quad \epsilon = \pm 1
$$
The Jacobian matrix of system (\ref{reduced}) at $V_\epsilon$ reads 
$$
A(V_\epsilon)= \left( 
  \begin{array}{cccc}
    - 2& 0 & 4 \epsilon & 0 \\
    0 & - \sigma \epsilon & 0 & - 2(\sigma \epsilon + 1)
    \\
    0& 0& 2& 0 \\
    0 & 0 & 0 & \sigma \epsilon
  \end{array}
\right). 
$$
The spectrum of the previous matrix is given by
$$
Sp(A(V_\epsilon)) = \{-2,\,\, -\sigma,\,\, \sigma,\,\, 2 \}.
$$
Each point $V_\epsilon$, $\epsilon = \pm 1$ has a stable manifold
$W_s^{(\epsilon)}$ and an unstable manifold $W_i^{(\epsilon)}$, both
of dimension two.

The points $P_{K^*}$ are {\sl attractors} (except for the marginal direction
connecting them), and our numerical simulations below will 
show that it is indeed possible to {\sl evolve from points $U_{K_*}$ at the
past light-cone to points $P_{K^*}$ at the future light-cone.} 

In this section, in order to establish 
that this is indeed the future light-cone $u=0$ of the singularity, 
we investigate the behavior of exterior incoming null rays for solutions
terminating at such point $P_{K^*}$. The exterior condition means that we
work with $\beta>0$. Evolving towards a fixed point means that we can
approach $x=+\infty$, and hence this is either $r=+\infty$ or $u=0$.
Therefore we need to show that incoming null rays can reach $x=+\infty$
at finite $r$. In self-similar coordinates $(\tau,s)$ the equation of
incoming null rays is
$$
\frac{dr}{du} = -\frac{1}{2}e^{\nu-\lambda},
\quad\Rightarrow\quad
\frac{ds}{d\tau} = \beta,
$$
and hence, for a given ray originating at $(\tau_0, s_0)$,
$$
\tau-\tau_0 = \int_{s_0}^s \alpha(s')ds'.
$$
From (\ref{expansionFLC}) we get that
$$
(\alpha(s)-1)e^{(1-{K^*}^2)s} \to c
\quad {\rm as} \quad
s \to +\infty
$$
for some constant $c$. Therefore, 
$$
\log r = s - \tau = s_0 - \tau_0 
-\int_{s_0}^s (\alpha(s')-1)ds'
$$
converges to a finite quantity as $s\to+\infty$.

Finally, it remains to show that {\sl the future light-cone is not a curvature
singularity, so that the spacetime can be continued beyond it.} 
Indeed, while the self-similar coordinate system $(\tau,x)$ becomes singular on the
future light-cone, we still can take the limit $x\to+\infty$ in the 
formula (\ref{RicciscalarB}), replacing the results in
(\ref{expansionFLC}). The result for the Ricci scalar $R$ of the spacetime
metric is
$$
r^2 R \to - \frac{2 {K^*}^2}{(1+{K^*}^2)({K^*}^2+\sigma^2)}
\quad {\rm as} \quad
x \to +\infty.
$$
Hence $R$ is {\sl finite everywhere on the future light-cone} ---except of course 
at $r=0$ (which is a curvature singularity). Similar expressions can be
derived for the Gauss curvature of the two-dimensional reduced spacetime, or for the
Kretchmann scalars of the four-dimensional and two-dimensional metrics.

Our points $P_{K^*}=(1,-K^*,0,K^*)$ play the same role as Christodoulou's
point $P_0=(1,-k)$, and we see that they are indeed closely related since 
the field $\xi$ vanishes at our points $P_{K^*}$. This can be
interpreted as a sign that the Brans-Dicke field is becoming {\sl irrelevant
on the future light-cone} of the singularity and, therefore, that the spacetime
has the same properties as the ones of Christodoulou's solutions within a small
neighborhood of the light-cone. 

An important difference between our construction and Christodoulou's
one is that he can construct time-symmetric solutions, since the
data on the past and future light-cones coincide. 
But, such construction is not possible here 
precisely because
 the point $U_{K_*}$ has a non-vanishing field
$\xi$ while the point $P_{K^*}$ has $\xi=0$. 

Christodoulou uses the above fact to ``copy'' the past region of the spacetime onto
the future region (cf. Fig.~\ref{regions}), finding a possible complete
spacetime containing the naked singularity, with a center which is 
only mildly singular. 
In our case, to have a complete spacetime, we would need to evolve (numerically) 
further from
data on the future light cone ---but we shall not do that here, as we decided 
focus on establishing the presence of the singularity, only.

Finally, the structure of phase space
around this point is quite different in our system. There is no equivalent
of Christodoulou's points $P_+$ and $P_-$, due to the general tendency of the
variables $\Theta$ and $\xi$ to 
{\sl ``rotate'' among them }
(which is the same phenomenon as the one discussed in Figure~\ref{streams}). The 
two-dimensional funnel in
Christodoulou's pictures is converted here into a {\sl higher-dimensional} 
analogue. In the following section we use numerical evolutions to demonstrate
this behavior.

%===================================================================================================

\section{Numerical investigations}
\label{section6}

\subsection{Numerical strategy}

We perform a numerical integration of the system~ (\ref{reduced}) using a
fourth-order Runge-Kutta scheme. This is done in four steps:
\begin{enumerate}
\item Using the coordinate $x=e^s$, from $x=0$, with
  $\alpha=0$, $\Theta=0$, $K=\kk$, and $\xi=0$, 
  we integrate the equation (\ref{399}) up to some
  finite value $x_0 = e^{s_0}$.
  
\item Then, switching to the reduced system (\ref{reduced}) in the coordinate $s$
  we use an adaptive-step approach to integrate on the interval $[s_0, s_*)$ and
   get
  as close as possible to $s_*$, where $\lim_{s\to s_*} \alpha = -\infty$.
  
\item Next, starting again from $s_1 > s_*$, we integrate the system (\ref{reduced}) 
  {\em backward\/} to reach $s_*$. At $s_1$, we have two new parameters,
  namely $\Theta_1 = \Theta(s_1)$ and $\xi_1 = \xi(s_1)$, whereas
  $\alpha(s_1)$ and $K(s_1)$ are determined so that all quantities $(\beta,
  \Theta, \xi, K)$ match at $s=s_*$.
  
\item Finally, from $s=s_1$, we integrate forward with a constant step-size and
  check whether the system (\ref{reduced}) converges to a stationary
  point or diverges.
\end{enumerate}
\noindent A first step is needed to start from the accurate values at the central
singularity, as defined in Section~\ref{ss:center}; the matching with the second
step at $s=s_0$ is straightforward, since the transition is done at a point at
which all quantities have regular behavior and we only make a change of
coordinate from $x$ to $s$. The matching at $s_*$ is much more complicated and
we first explain now the technique used to recover the results by
Christodoulou~\cite{Christodoulou94}. 

\subsection{Numerical integration of Christodoulou's system}

In his study, Christodoulou solves the
equivalent of our two first equations (for $\alpha=\alpha(s)$ and $\Theta=\Theta(s)$) in the
reduced system (\ref{reduced}):
\begin{eqnarray}
  \label{e:christo_alpha}
  \frac{\textrm{d}\alpha}{\textrm{d}s} &={} \alpha \left( \left( \theta + k
    \right)^2 + \left( 1 - k^2 \right)\left( 1 - \alpha \right) \right),
    \\
  \frac{\textrm{d}\theta}{\textrm{d}s} &={} k\alpha \left( k\theta - 1 \right)
  + \theta \left( \left( \theta + k \right)^2 - \left( 1 + k^2 \right)
  \right).   \label{e:christo_theta}
\end{eqnarray}
(Cf.~the equations $(1.1a)$ and $(1.1b)$ in \cite{Christodoulou94}.)
In a neighborhood of $s_*$, for $s>s_*$, Christodoulou finds that the solution
$\theta$ depends on a real parameter $a_1$ (see~(2.13)
and~(2.14) in~\cite{Christodoulou94}), but one always has
$$
\lim_{s\to s_*} \theta = \frac{1}{k}.
$$
On the other hand, the solution $\beta = \alpha^{-1}$ has the following
behavior:
\begin{equation}
  \label{e:local_beta_christo}
  \beta = (1 - k^2) \left( s - s_* \right) + \mathcal{O} \left( \left| s - s_*
    \right|^2 \right).
\end{equation}
With these information, we devise the numerical integration strategy as
follows. Given a value of $k$, we integrate the
system (\ref{e:christo_alpha})--(\ref{e:christo_theta}) until $s\to s_*$, and
thus determine the approximate value of $s_*$ up to high accuracy. We then
define 
\begin{equation}
  \label{e:def_s1}
  s_1 = (1 \pm \epsilon) s_*,
\end{equation}
with the sign chosen so as to $s_1 > s_*$ and $\epsilon \sim 0.03$ for
numerical convenience. We then set
\begin{eqnarray}
  \label{e:def_beta1_theta1}
  \alpha(s_1) &={} \frac{1}{(1 - k^2)\epsilon s_*},\\
  \theta(s_1) &={} \theta_1,
\end{eqnarray}
with $\theta_1$ a new parameter than can be freely chosen and is to represent
the degree of freedom, induced by the parameter $a_1$ of Christodoulou's study
(see Section~2 and Eqs.~(2.13b)-(2.14c) of \cite{Christodoulou94}). From the two
points $(s_1, \alpha_1), (s_1, \theta_1)$, we integrate backward toward $s_*$
and $\alpha(s) \to +\infty$. When doing so, we find numerically that, in most cases, 
$\alpha$ is diverging at some value $s'_* \not= s_*$. This is due
to the approximate value of $\alpha_1$ in (\ref{e:def_beta1_theta1}) that was
chosen to initiate the integration. Since we are dealing with an autonomous system,
we can perform a slight shift $\Delta s = s'_* - s_*$ in the variable $s$, so
that $\lim_{s\to s_*^+} \alpha(s) = +\infty$.
\begin{figure}
  \centerline{\includegraphics[width = 10cm]{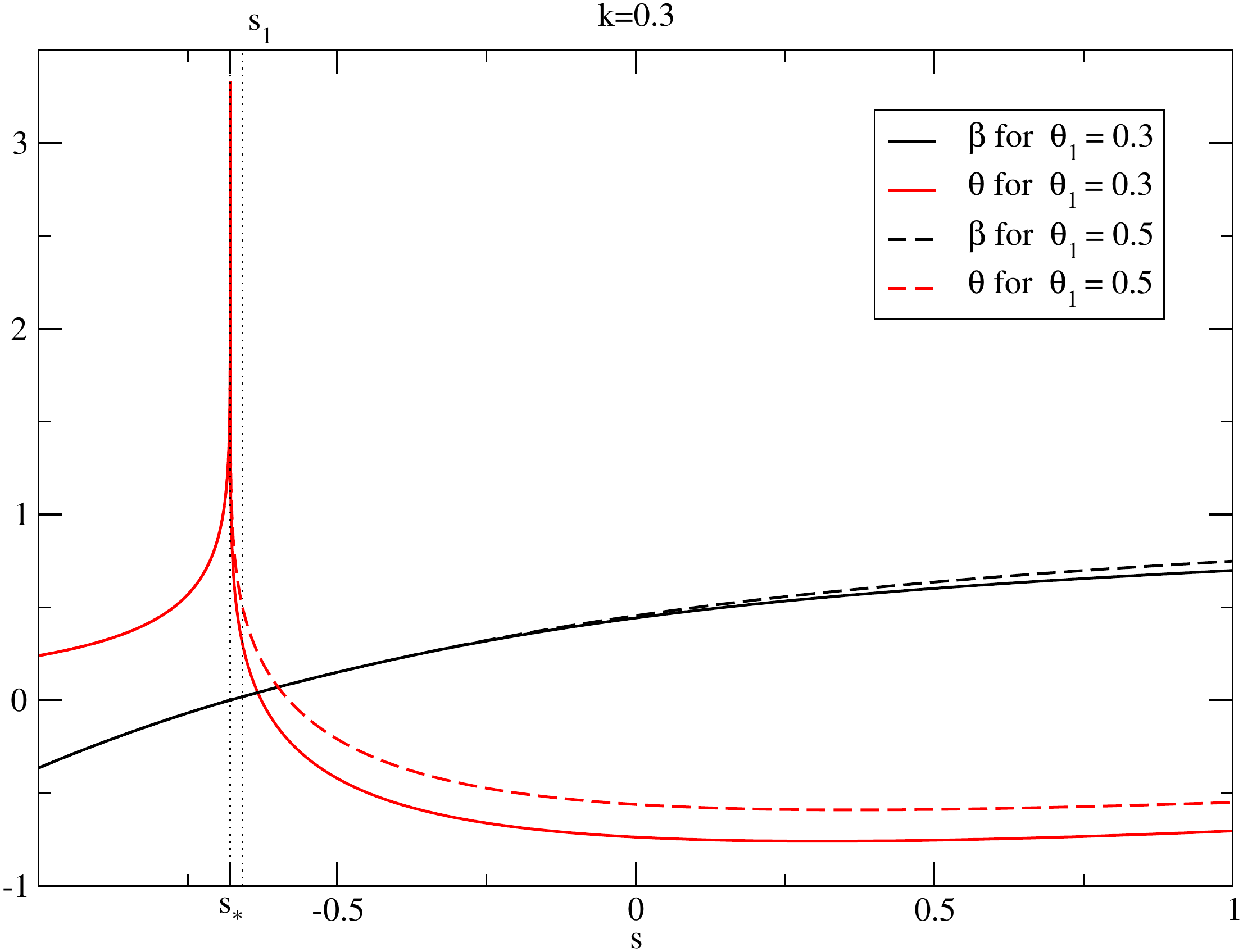}}
  \caption{Solutions $\beta(s)=\alpha(s)^{-1}$ and $\theta(s)$ in the case of
    Christodoulou's system (\ref{e:christo_alpha})--(\ref{e:christo_theta}).
    The
    matching of the solution at $s=s_*$ is shown for two different values of the
    parameter $\theta_1 = \theta(s_1)$.}
  \label{f:christo}
\end{figure}

In Figure~\ref{f:christo} are shown the numerical solutions of the differential
system (\ref{e:christo_alpha}--(\ref{e:christo_theta}), for $\kk=0.3$ and with
two different values of $\theta_1 = \theta(s_1)$. We have numerically observed
that, if the parameter $\epsilon$ was small enough and for any value of
$\theta_1$, the backward integration detailed here-above would always bring
back to the solution $\beta(s_*) = 0$ and $\theta(s_*) = 1/k$. We were thus
able to numerically recover the result by
Christodoulou~\cite{Christodoulou94} that, for a given $k<1$, each solution of
the system (\ref{e:christo_alpha})--(\ref{e:christo_theta}) connects to a
one-parameter family of solutions at $s=s_*$. 

\begin{figure}
  \centerline{\includegraphics[width = 10cm]{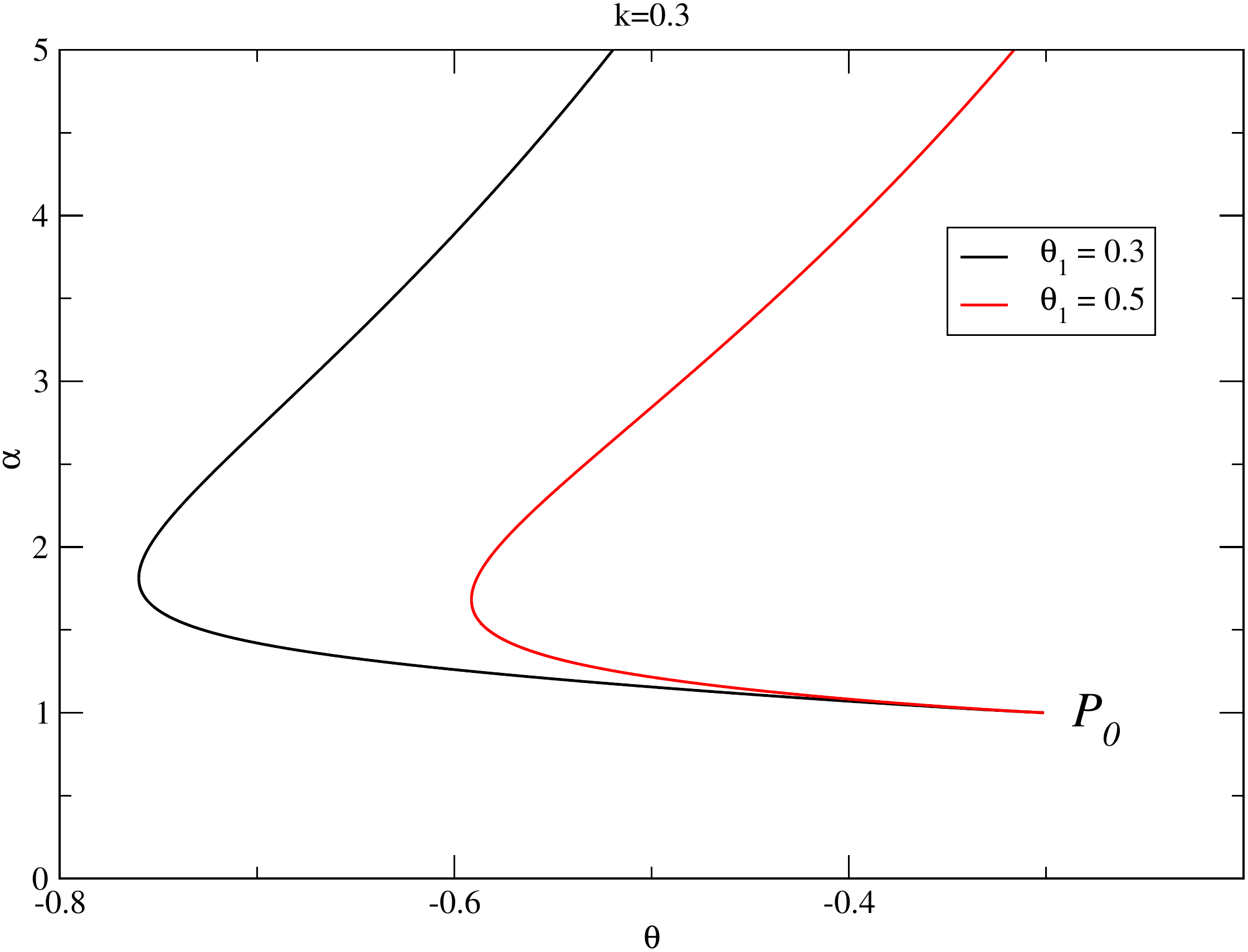}}
  \caption{Trajectories in $(\theta,\alpha)$ plane of the two solutions shown
    in Figure~\ref{f:christo}. In both cases, the stationary point
    $\mathcal{P}_0:(1,-k)$ is reached as $s\to +\infty$.}
  \label{f:alpha_christo}
\end{figure}

By varying the parameter $\theta_1$, one can reach different asymptotic
regimes, when $s\to +\infty$: both fields $\alpha$ and $\theta$ can diverge or
they can converge to the stationary point $\mathcal{P}_0 : (\alpha=1, \theta =
-k)$. This last case is displayed in Figure~\ref{f:alpha_christo}, with the
trajectories of the solutions in the $(\theta,\alpha)$ plane. 
(This is to be
compared with Figure~4 in~\cite{Christodoulou94}.)

\subsection{Numerical solutions of the reduced system}\label{ss:num_sol}

In the case of interest in this paper, we have to deal with the matching of
four fields $(\alpha, \Theta, \xi, K)$ at $s=s_*$. We use the same numerical
technique, with the four steps described at the beginning of this section. We
have numerically observed that here, in addition to $\theta_1$($\Theta_1 =
\Theta(s_1)$), we need to specify the value of $\xi_1 = \xi(s_1)$. On the
other hand, near an equilibrium point $U_{K_*}$, with $ 0 < K_* <1$ we get
from the first and last equations in (\ref{reducedbeta}): 
$$
\beta'(s) \sim 1 - K_*^2, \qquad \quad  K'(s) \sim \sigma K_* \xi_*,
$$
so that  we can write the following expansions: 
\begin{eqnarray}
  \beta(s) &={} \left( 1 - K_*^2 \right) \left( s - s_* \right) + o
  \left( \left| s - s_* \right| \right) \label{e:dl_beta},\\
  K(s) &={} K_* - \frac{\sigma^2 K_*}{K_*^2 + \sigma^2} \left( s - s_* \right) +
  o \left( \left| s - s_* \right| \right) \label{e:dl_K}.
\end{eqnarray}
We can therefore obtain numerical estimations of the values of these two
fields at $s=s_1$ and perform the integration backward, from $s_1$ toward
$s_*$. We again do the shift in $s$ to match $\beta$ up to machine precision
at $s=s_*$, but doing so does not allow for an accurate matching of $K(s)$. We
then do several (usually no more than five) integrations from $s_1$ toward
$s_*$, correcting each time the starting value $K(s_1)$ in such a way that, at
the end, the function $K(s)$ is continuous at $s_*$, up to machine precision.  

\begin{figure}
  \centerline{\includegraphics[width = 10cm]{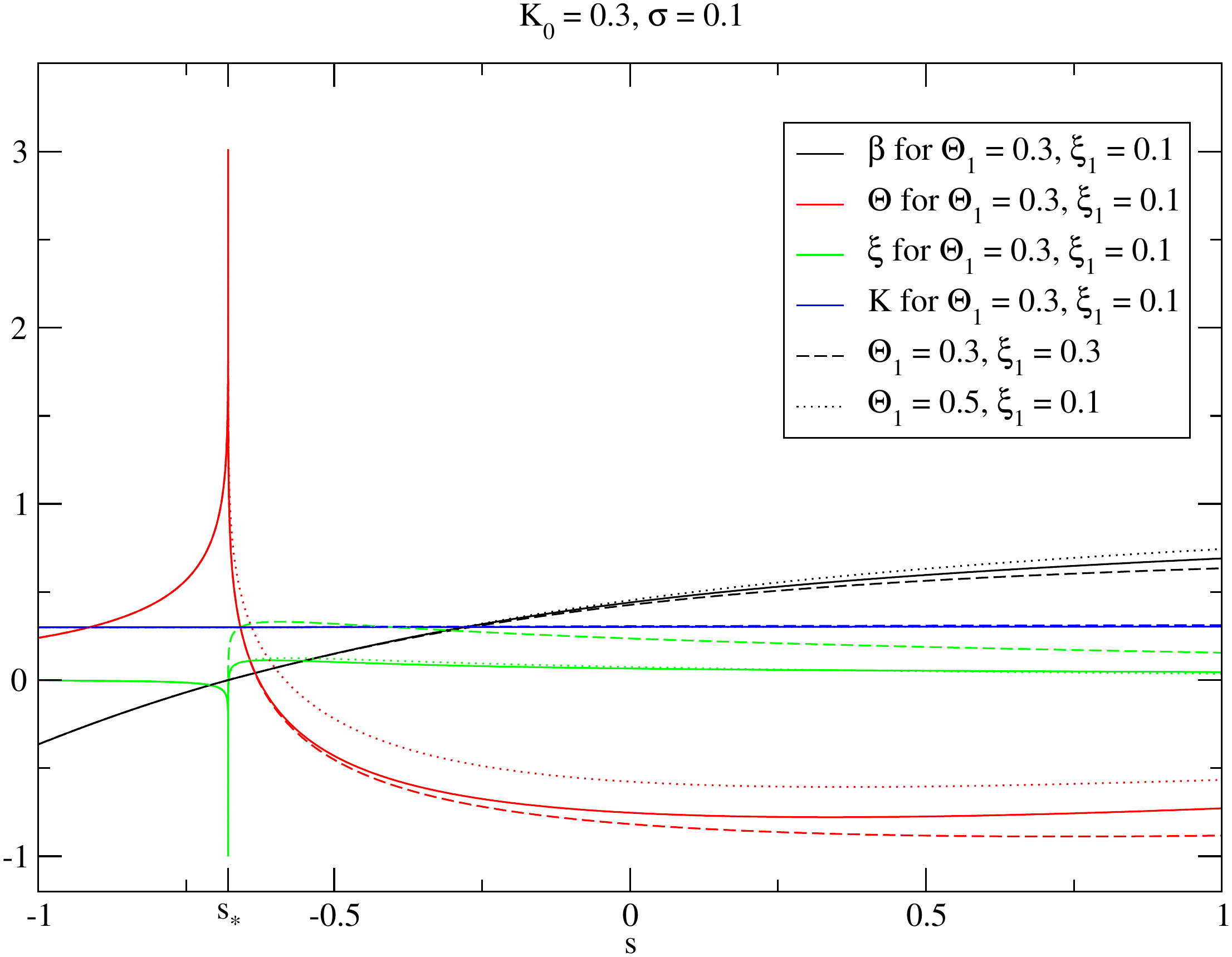}}
  \caption{Solutions $\beta \left(= \alpha^{-1}\right), \Theta, \xi$ and $K$
    of the reduced system (\ref{reduced}). The matching of the fields at
    $s=s_*$ are shown for three sets of the parameters $\left(\Theta_1=
    \Theta(s_1), \xi_1 = \xi(s_1)\right)$.}
  \label{f:resu_tss}
\end{figure}

Results are displayed in Figure~\ref{f:resu_tss}, where we have taken $\kk=0.3$
and $\sigma = 0.1$. The matching of all four fields are done with the setting
of two new parameters $\left(\Theta_1, \xi_1 \right)$, which seems to indicate
that a solution starting from the interior region ($s<s_*$) connects to a
two-parameter family of solutions in the exterior region ($s>s_*$). Depending
on these two parameters $\left(\Theta_1, \xi_1 \right)$, we numerically
recover a behavior similar to that of Christodoulou's
system~\cite{Christodoulou94}: for some values of $k, \sigma, \Theta_1$ and
$\xi_1$, the system can converge to the stationary point (b) of
Section~\ref{ss:center}, that is, $\left(1, -K^*, 0, K^* \right)$ with 
\begin{equation}
  \label{e:Klim}
  K^* = \lim_{s\to+\infty} K(s).
\end{equation}

\begin{figure}
  \centerline{\includegraphics[width = 10cm]{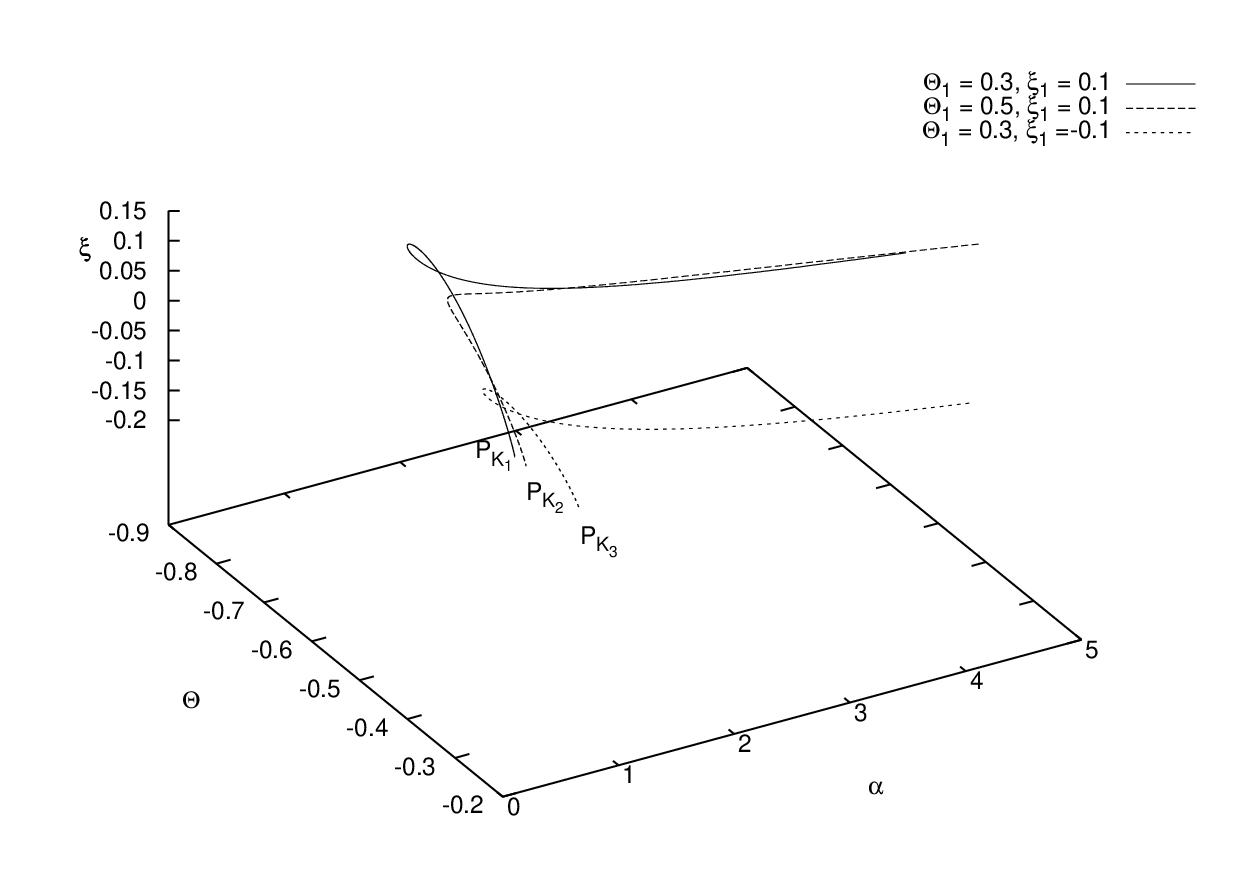}}
  \caption{Trajectories of the solutions of the reduced system (\ref{reduced})
    in the $(\theta,\alpha,\xi)$ space, for three different sets of parameters
    $\left(\Theta_1, \xi_1\right)$.}
  \label{f:alpha_tss}
\end{figure}

Part of this behavior is displayed in Figure~\ref{f:alpha_tss}, where the
trajectories in the $\left( \theta, \alpha, \xi \right)$ space, for three sets
of parameters $\left(\Theta_1, \xi_1\right)$. Each set of parameters can lead
\textit{a priori} to a different limit $P_{K^*}$. If these trajectories were
projected onto the $(\alpha, \Theta)$ plane, they would resemble a lot the
ones of the general-relativistic system of Figure~\ref{f:alpha_christo}, studied
by Christodoulou~\cite{Christodoulou94}, with the noticeable difference that we
no longer have a single limit $\mathcal{P}_0$, but the endpoint depends in
general on the value $K^*$ (see~(\ref{e:Klim})), which changes from one set
of parameters to another.

%==============================================================================================

\section{Conclusions}
\label{section7}

We have studied the formation of naked singularities in the process of the gravitational collapse of a real massless scalar field, and have generalized Christodoulou's construction of a family of spacetimes 
containing a naked singularity. 
While Christodoulou worked within the classical Einstein theory, we have here considered 
the Brans-Dicke theory. This effectively led us to deal with 
a new scalar field and to promote Christodoulou's constant 
parameter $k$ to a dynamical variable $K$.

We were able to fully analyze the interior region and rigorously justify the matching across the past light-cone of the naked singularity. Partial analytical information was also obtained in the exterior region, and 
we finally completed our study with numerical simulations. We could 
show that the variable $K$ always decreases from any initial value to a value $K_*$
smaller than $1$ at the past light-cone. 
This eliminates the possibility in Christodoulou's work of forming
pathological solutions without a past light-cone. In that sense the
addition of the Brans-Dicke field has a ``regularizing'' effect.
The behavior of the solutions in the exterior region is 
similar to the one of Christodoulou's solution and, in fact, the Brans-Dicke
field vanishes on the future light-cone of the singularity (a Cauchy
horizon). However, the dynamical structure of the phase space is quite different,
as our phase space is much larger and does not contain Christodoulou's case as a subspace.

Like Christodoulou established for solutions to the classical Einstein equations, 
our solutions are probably highly unstable against small (radially symmetric, for instance) perturbations.

It is quite reasonable to expect that more general scalar-tensor theories
would exhibit a similar behavior. In particular, it would be interesting to extend our conclusions
 to the more general models arising in 
the so-called $f(R)$ theories of gravity when the action involves a nonlinear function of the Ricci scalar.

\bigskip
\noindent {\bf Acknowledgments}

The authors were supported by the Agence Nationale de la Recherche (ANR) through the grant 06-2-134423 entitled
``Mathematical Methods in General Relativity'' (Math-GR).
JMM was also supported by the French ANR grant BLAN07-1\_201699 entitled
``LISA Science'', and also in part by the Spanish MICINN projects
FIS2009-11893 and FIS2008-06078-C03-03. 
PLF and JMM acknowledge support from the Erwin Schr\"odinger Institute,
Vienna, during the program {\sl ``Quantitative Studies of Nonlinear Wave Phenomena'',}  
organized by P.C. Aichelburg, P. Bizo\'n, and W. Schlag.

%===================================================================================================

\section*{References}

\bibliographystyle{prsty}

\end{document}